\documentclass[twocolumn,prl,aps,epsfig]{revtex4}
\usepackage{epsfig}
\def\beq{\begin{equation}}
\def\enq{\end{equation}}
\def\beqa{\begin{eqnarray}}
\def\enqa{\end{eqnarray}}
\def\MeV{\nobreak\,\mbox{MeV}}
\def\GeV{\nobreak\,\mbox{GeV}}

\def\pli{p^\prime}

\def\La{\Lambda}

\def\la{\lambda}

\def\rh{\rho}
\def\si{\sigma}

\def\al{\alpha}

\def\qq{\lag\bar{q}q\rag}
\def\ss{\lag\bar{s}s\rag}
\def\mix{\lag\bar{q}g\si.Gq\rag}
\def\mixs{\lag\bar{s}g\si.Gs\rag}
\def\Gd{\lag g^2G^2\rag}
\def\G3{\lag g^3G^3\rag}
\def\lb{\label}
\def\nn{\nonumber}
\def\aa{\scriptsize\mbox{a }}
\def\bb{\scriptsize\mbox{b }}

\newcommand{\rag}{\rangle}
\newcommand{\lag}{\langle}

\begin{document}
\title{Diquark-Antidiquark with open charm in QCD sum
rules}
\author{M. Nielsen $^{\aa}$, R.D. Matheus $^{\aa}$, F.S. Navarra$^{\aa}$,
M.E. Bracco $^{\bb}$ and A. Lozea $^{\bb}$}
\affiliation{$^{\aa}$Instituto de F\'{\i}sica, 
        Universidade de S\~{a}o Paulo, 
        C.P. 66318,  05389-970 S\~{a}o Paulo, SP, Brazil\\
        $^{\bb}$ Instituto de F\'{\i}sica, Universidade do Estado do Rio de 
Janeiro, 
Rua S\~ao Francisco Xavier 524, 20550-900 Rio de Janeiro, RJ, Brazil}

\begin{abstract}
Using the QCD sum rule approach we investigate the 
possible four-quark structure of the recently observed charmed scalar mesons 
$D_0^{0}(2308)$ (BELLE) and $D_0^{0,+}(2405)$ (FOCUS) and also of the very 
narrow $D_{sJ}^{+}(2317)$, firstly observed by BABAR. 
We use  diquak-antidiquark 
currents and work to the order of $m_s$ in full QCD, without relying on 
$1/m_c$ expansion. Our results indicate that a four-quark structure is 
acceptable for the resonances observed by BELLE and BABAR: $D_0^{0}(2308)$ 
and $D_{sJ}^{+}(2317)$ respectively, but not for the resonances observed 
by FOCUS: $D_0^{0,+}(2405)$.  
\end{abstract}

\maketitle

In general, the classification of mesons containing a single heavy quark is
interpreted with the help of heavy-quark symmetry, {\it i.e.}, the symmetry
valid for the infinitely heavy mass of charm quark. Under this symmetry, the
strong interaction conserves total angular momentum of the light quark, $j$. 
In the meanwhile, total angular momentum of the light-heavy system, $J$,
should be still  regarded as a good quantum number of the system, even if the
 heavy-quark symmetry breaks down. In this way, the classification of the
charmed mesons can be explained in terms of the quantum numbers $(L,S,J,j)$,
where $L$ and $S$ denote the orbital angular momentum between the light and
heavy quarks and total spin of the system, respectively. The doublets with
$j=1/2$ and $L=0$ have been observed over the past two decades ($\sim1975~-
~1994)$ following the discovery of open charm, because these states have 
relatively narrow widths. Recently the first observations of the scalar 
charmed mesons $(j=1/2,~L=1)$ have been 
reported. The very narrow $D_{sJ}^+(2317)$ was first discovered in the
$D_s^+\pi^0$ channel by the BABAR Collaboration \cite{babar} and its
existence was confirmed by CLEO \cite{cleo}, BELLE \cite{belle1} and
FOCUS \cite{focus} Collaborations. Its mass was commonly measured as
$2317 \MeV$, which is approximately $160 \MeV$ below the prediction of 
the very successful quark model for the charmed mesons \cite{god}. 
The BELLE Collaboration \cite{belle2} has also reported the observation of 
a rather broad scalar meson $D_0^{0}(2308)$, and the FOCUS Collaboration
\cite{focus2} reported evidence for broad structures in both neutral and
charged final states that, if interpreted as resonances in the $J^P=0^+$
channel, would be  the $D_0^{0}(2407)$ and the $D_0^{+}(2403)$ mesons. 
While the 
mass of the scalar meson, $D_0^{0}(2308)$, observed by  BELLE Collaboration
is also bellow the prediction of ref.~\cite{god} (approximately $100 \MeV$),
the masses of the states observed by FOCUS Collaboration are in complete
agreement with ref.~\cite{god}.

\begin{figure} \label{fig1}
\centerline{\epsfig{figure=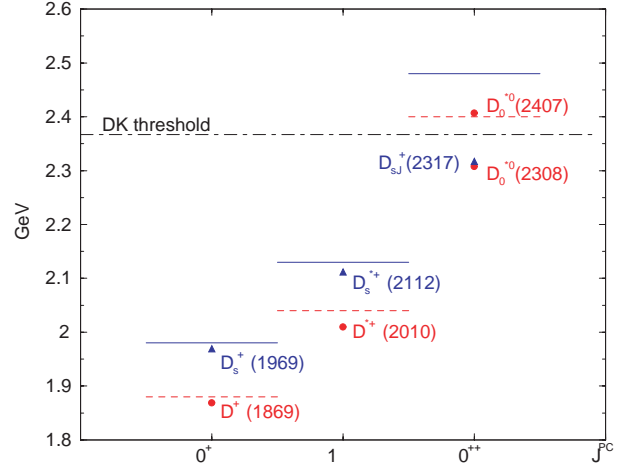,width=8cm}}
%\vspace{-1cm}
\caption{\small{Spectroscopy of the pseudoscalar, vector and scalar charmed 
mesons.The theoretical predictions of ref.~\protect{\cite{god}} are 
represented as solid lines for
the $c\bar{s}$ and dashed lines for the $c\bar{q}$, and the experimental
data are represented as  triangles for
the $c\bar{s}$ and circles for the $c\bar{q}$.}}  
\end{figure}

The spectroscopy of $c\bar{q}$ and $c\bar{s}$
pseudoscalar, vector and scalar mesons is drawn in Fig.~1, where the 
theoretical predictions of ref.~\cite{god} are represented as solid lines for
the $c\bar{s}$ and dashed lines for the $c\bar{q}$, and the experimental
data are represented as  triangles for
the $c\bar{s}$ and circles for the $c\bar{q}$.

Due to its low mass, the structure of the meson $D_{sJ}^+(2317)$ has been 
extensively debated. It has been interpreted as 
a $c\bar{s}$ state \cite{dhlz,bali,ukqcd,ht,nari}, two-meson molecular 
state \cite{bcl,szc}, $D-K$- mixing \cite{br},  
four-quark states \cite{ch,tera,mppr} or a mixture between two-meson
and four-quark states \cite{bpp}. The same analyses would also apply
to the meson $D_0^{0}(2308)$.

In the light sector the idea that the light
scalar mesons (the isoscalars $\sigma(500),~f_0(980)$, the isodublet 
$\kappa(800)$ \cite{E791} and the isovector $a_0(980)$) 
could be four-quark bound states is not new \cite{jaffe,cloto}. Indeed, in a
four-quark scenario, the mass degeneracy of $f_0(980)$ and $a_0(980)$ is 
natural, the mass hierarchy pattern of the nonet is understandable, and
it is easy to explain why $\sigma$ and $\kappa$ are broader than $f_0(980)$ 
and $a_0(980)$. The decays $\sigma\to\pi\pi$,
$\kappa\to K\pi$ and $f_0,~a_0\to KK$ are OZI superallowed without the need 
of any gluon exchange, while $f_0\to\pi\pi$ and $a_0\to\eta\pi$ are OZI
allowed as it is mediated by one gluon exchange. Since $f_0(980)$ and 
$a_0(980)$ are very close to the $\bar{K}K$ threshold, the $f_0(980)$
is dominated by the $\pi\pi$ state and $a_0(980)$ is governed by
the $\eta\pi$ state. Consequently, they are narrower than $\sigma$
and $\kappa$. 

In this work we use the  method of QCD  sum rules (QCDSR) \cite{svz}
to study the two-point functions of the scalar mesons, $D_{sJ}(2317)$, 
$D_0(2308)$ and $D_0(2405)$ considered as four-quark states. 
In a recent calculation \cite{sca}  the light scalar mesons were
considered as $S$-wave bound states of a diquark-antidiquark pair. As 
suggested in ref.~\cite{jawil} the diquark was taken to be a spin zero 
colour anti-triplet. We extend this prescription
to the charm sector and, therefore, the corresponding interpolating fields 
containing zero, one and two strange quarks are:
\beqa
j_0&=&\epsilon_{abc}\epsilon_{dec}(q_a^TC\gamma_5c_b)
(\bar{u}_d\gamma_5C\bar{d}_e^T),
\nn\\
j_1&=&{\epsilon_{abc}\epsilon_{dec}\over\sqrt{2}}\left[(u_a^TC
\gamma_5c_b)(\bar{u}_d\gamma_5C\bar{s}_e^T)+u\leftrightarrow d\right],
\nn\\
j_{2}&=&\epsilon_{abc}\epsilon_{dec}(s_a^TC
\gamma_5c_b)(\bar{q}_d\gamma_5C\bar{s}_e^T),
\label{int}
\enqa
where $q$ represents the quark $u$ or $d$ according to the charge of the meson.
Since $D_{sJ}$ has one $\bar{s}$ quark, we choose the $j_1$ current to 
have the same quantum numbers of $D_{sJ}$, which is supposed to be
an isoscalar. However, since we are working in the SU(2) limit, the
isoscalar and isovector states are mass degenerate and, therefore, this
particular choice has no relevance here.

The QCDSR for the charmed scalar mesons with $n$ strange quarks are 
constructed from the two-point correlation function
\beq
\Pi(q)=i\int d^4x ~e^{iq.x}\lag 0 |T[j_n(x)j^\dagger_n(0)]|0\rag.
\lb{2po}
\enq

In the OPE side we work at leading order and consider condensates up to 
dimension six. We deal with the strange quark as a light one and consider
the diagrams up to order $m_s$. To keep the charm quark mass finite, we
use the momentum-space expression for the charm quark propagator. We 
calculate the light quark part of the correlation
function in the coordinate-space, which is then Fourier transformed to the
momentum space in $D$ dimensions. The resulting light-quark part is combined 
with the charm-quark part before it is dimensionally regularized at $D=4$.

We can write the correlation function in the OPE side in terms of a 
dispersion relation:
\beq
\Pi^{OPE}(q^2)=\int_{m_c^2}^\infty ds {\rho(s)\over s-q^2}\;,
\lb{ope}
\enq
where the spectral density is given by the imaginary part of the correlation
function: $\rho(s)={1\over\pi}\mbox{Im}[\Pi^{OPE}(s)]$.

In the phenomenological side, the coupling of the scalar meson with $n$ strange 
quarks, $S_n$, to the scalar current, $j_n$,  can be
parametrized in terms of the meson decay constant $f_{S_n}$ as \cite{sca}:
$\lag 0 | j_n|S_n\rag =\sqrt{2}f_{S_n}m_{S_n}^4$,
therefore, the phenomenological side of Eq.~(\ref{2po}) can be written as
\beq
\Pi^{phen}(q^2)={2f_{S_n}^2m_{S_n}^8\over m_{S_n}^2-q^2}+\cdots\;,
\lb{phe}
\enq
where the dots denote higher resonance contributions that will be 
parametrized, as usual, through the introduction of the continuum threshold
parameter $s_0$ \cite{io1}.  After making a Borel
transform on both sides, and transferring the continuum contribution to
the OPE side, the sum rule for the scalar meson $S_n$ can be written as
\beq
2f_{S_n}^2m_{S_n}^8e^{-m_{S_n}^2/M^2}=\int_{m_c^2}^{s_0}ds~ e^{-s/M^2}~
\rho_{S_n}
(s)\;,
\lb{sr}
\enq
where $\rho_{S_n}(s)=\rho^{pert}(s)+\rh^{m_s}(s)+\rh^{\qq}(s)+\rh^{\lag G^2
\rag}
(s)+\rh^{mix}(s)+\rh^{\qq^2}(s)+\rh^{\lag G^3\rag}(s)$, with \cite{blmnn}
\beq
\rho^{pert}(s)={1\over 2^{10} 3\pi^6}\int_\La^1 d\al\left({1-\al\over\al}
\right)^3(m_c^2-s\al)^4,
\enq
\beqa
\rho^{\lag G^2\rag}(s)={\Gd\over 2^{10}\pi^6}\int_\La^1 d\al~(m_c^2-s\al)
\bigg[{m_c^2\over9}
\nn\\
\times\left({1-\al\over\al}\right)^3+
(m_c^2-s\al)\left({1-\al\over2\al}+{(1-\al)^2\over4\al^2}\right)
\bigg],
\enqa
\beq
\rho^{\lag G^3\rag}(s)={\G3\over 2^{12} 9\pi^6}\int_\La^1 d\al\left({1-\al
\over\al}\right)^3(3m_c^2-s\al),
\enq
which are common to all three resonances and where the lower limit of the 
integrations is given by $\La=m_c^2/s$. From $j_0$ we get: $\rh^{m_s}(s)=0$,
\beq
\rho^{\qq}(s)=-{m_c\qq\over 2^{6}\pi^4}\int_\La^1 d\al\left({1-\al\over\al}
\right)^2(m_c^2-s\al)^2,
\enq
\beqa
&&\rho^{mix}(s)={m_c\mix\over 2^{6}\pi^4}\bigg[{1\over2}\int_\La^1 d\al
\left({1-\al\over\al}\right)^2
\nn\\
&\times&(m_c^2-s\al)-\int_\La^1 d\al{1-\al\over\al}(m_c^2-s\al)\bigg],
\enqa
\beq
\rho^{\qq^2}(s)=-{\qq^2\over 12\pi^2}\int_\La^1 d\al~(m_c^2-s\al).
\enq
From $j_{1}$ we get: $\rh^{m_s}(s)=0$,
\beqa
\rho^{\qq}(s)={1\over 2^{6}\pi^4}\int_\La^1 d\al~{1-\al\over\al}
(m_c^2-s\al)^2
\bigg[
\nn\\
-\qq\left(2m_s+m_c{1-\al\over\al}\right)+m_s\ss\bigg],
\enqa
\beqa
\rho^{mix}(s)&=&{1\over 2^{6}\pi^4}\int_\La^1 d\al~(m_c^2-s\al)\bigg[-{m_s\mixs
\over6}
\nn\\
&+&\mix\bigg(-m_s(1-\ln(1-\al))
\nn\\
&-&m_c{1-\al\over\al}\left(1-{1\over2}{1-\al\over\al}\right)\bigg)
\bigg]
\enqa
\beq
\rho^{\qq^2}(s)=-{\qq\ss\over 12\pi^2}\int_\La^1 d\al~(m_c^2-s\al).
\enq
Finally from $j_{2}$ we get
\beq
\rh^{m_s}(s)=-{m_sm_c\over 2^{8} 3\pi^6}\int_\La^1 d\al\left({1-\al\over\al}
\right)^3(m_c^2-s\al)^3,
\enq
\beqa
\rho^{\qq}(s)={1\over 2^{6}\pi^4}\int_\La^1 d\al{1-\al\over\al}(m_c^2-s\al)^2
\bigg[
\nn\\
\ss\left(2m_s-m_c{1-\al\over\al}\right)-2m_s\qq\bigg],
\enqa
\beqa
&&\rho^{mix}(s)={1\over 2^{6}\pi^4}\int_\La^1 d\al~(m_c^2-s\al)\bigg[{\mixs
\over2}
\nn\\
&\times&\bigg({m_s\over3}-m_s{1-\al\over\al}
-m_c{1-\al\over\al}\left(1-{1\over2}{1-\al\over\al}\right)\bigg)
\nn\\
&-&m_s\mix(1-\ln(1-\al))\bigg],
\enqa
\beq
\rho^{\qq^2}(s)=-{\qq\ss\over 12\pi^2}\int_\La^1 d\al~(m_c^2-s\al).
\enq

In the numerical analysis of the sum rules, the values used for the quark
masses and condensates are: $m_s=0.13\,\GeV$, $m_c=1.2\,\GeV$, 
$\lag\bar{q}q\rag=\,-(0.23)^3\,\GeV^3$,
$\langle\overline{s}s\rangle\,=0.8\lag\bar{q}q\rag$,  
$\lag\bar{q}g\si.Gq\rag=m_0^2
\lag\bar{q}q\rag$ with $m_0^2=0.8\,\GeV^2$, $\lag g^2G^2\rag=0.5~\GeV^4$
and $\lag g^3G^3\rag=0.045~\GeV^6$.

We call $D_0^{(0s)}$, $D_{0}^{(1s)}$ and $D_{0}^{(2s)}$ the scalar charmed
 mesons represented by $j_0$,
$j_1$ and $j_{2}$ (in Eq.~(\ref{int})) respectively.
\begin{figure}[htb]
%\vspace{9pt}
\centerline{\epsfig{figure=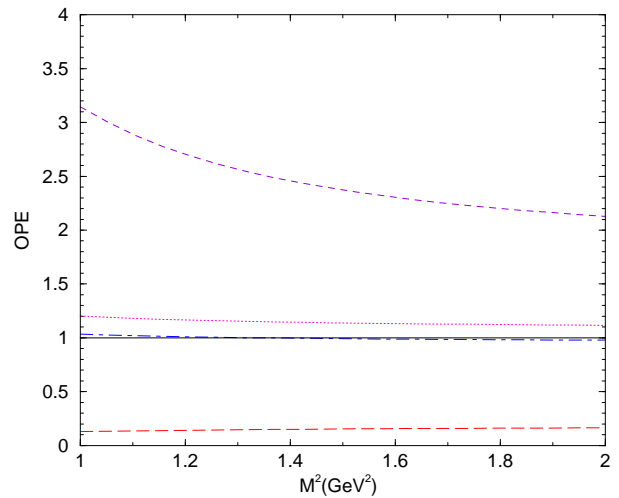,width=8cm}}
%\vspace{-1cm}
\caption{\small{Borel mass dependence of the relative contributions of the OPE
terms: perturbative (long-dashed 
line), perturbative plus quark condensate (dot-dashed line),  previous
plus four-quark condensate (dashed line), previous plus mixed condensate 
(solid line).}}
\label{fig2}
\end{figure}
In Fig. 2 we show, as a function of the Borel mass, the OPE relative 
contribution of the: perturbative (long-dashed 
line), perturbative plus quark condensate (dot-dashed line),  previous
plus four-quark condensate (dashed line) and previous plus mixed condensate 
(solid line), for the $D_{0}^{(1s)}$ meson. We see that there is no good OPE 
convergence and that the four-quark condensate and the mixed condensate 
contributions are very big, as compared with the perturbative contribution, 
and with opposite signal, in such a way that the final 
result is almost the same as before adding these two contributions. One can 
argue that this is not a good sum rule, since there is not a good OPE 
convergence.
We notice, however, that this is a common feature of the sum rules for currents
with more than three quarks. The thre-gluon
condensate contribution is negligible for all three currents and this is why we
do not show it in Fig.~2. The same behaviour is obtained for $D_{0}^{(0s)}$
and $D_{0}^{(2s)}$ mesons.

\begin{figure}[htb]
%\vspace{9pt}
\centerline{\epsfig{figure=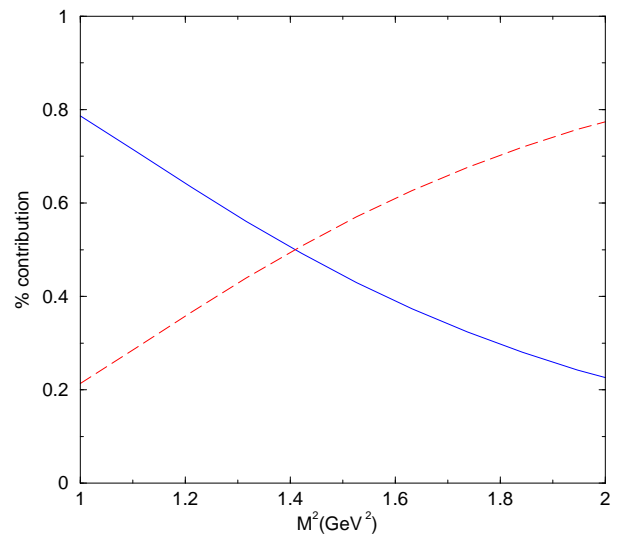,width=8cm}}
%\vspace{-1cm}
\caption{\small{Borel mass dependence of the relative contributions of the
pole (solid line) and continuum (dashed line) contributions.}}
\label{fig3}
\end{figure}

In Fig. 3 we show, as a function of the Borel mass, the percentage
of the pole and continuum contributions to the total contribution  for the 
$D_{0}^{(1s)}$ meson. We see that
in the Borel window $1.0\GeV^2\leq M^2\leq 1.4\GeV^2$ the pole contribution is
always bigger than the continuum contribution. Therefore, this is the Borel
window that we will consider.
\begin{figure}[h]
%\vspace{9pt}
\centerline{\epsfig{figure=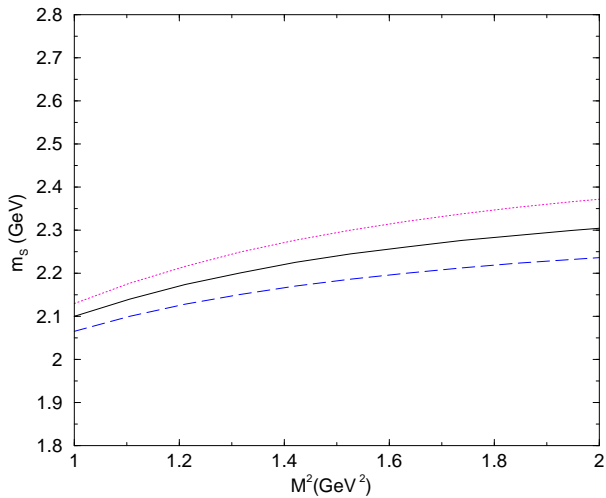,width=8cm}}
%\vspace{-1cm}
\caption{\small{The $D_{0}^{(0s)}$ mass as a function of the Borel mass 
for different
values of the continuum threshold.  Dashed line: $\sqrt{s_0}=2.6\GeV$; solid
line: $\sqrt{s_0}=2.7\GeV$; dotted line: $\sqrt{s_0}=2.8\GeV$.}}
\label{fig4}
\end{figure}
In order to get rid of the meson decay constant and
extract the resonance mass, $m_{S_n}$, we first take the derivative
of Eq.~(\ref{sr}) with respect to $1/M^2$ and then we divide it by
Eq.~(\ref{sr}) to get
\beq
m_{S_n}^2={\int_{m_c^2}^{s_0}ds ~e^{-s/M^2}~s~\rh_{S_n}(s)\over
\int_{m_c^2}^{s_0}ds ~e^{-s/M^2}~\rh_{S_n}(s)}\;.
\lb{m2}
\enq
In Figs. 4 and 5 we show the masses of the $D_0^{(0s)}$ and $D_0^{(1s)}$ 
resonances, respectively,
as a function of the Borel mass for different values of the continuum 
threshold. The results for the $D_0^{(2s)}$ resonance is similar to that
for the $D_0^{(1s)}$ resonance, as shown in ref.\cite{blmnn}.
\begin{figure}[h]
%\vspace{9pt}
\centerline{\epsfig{figure=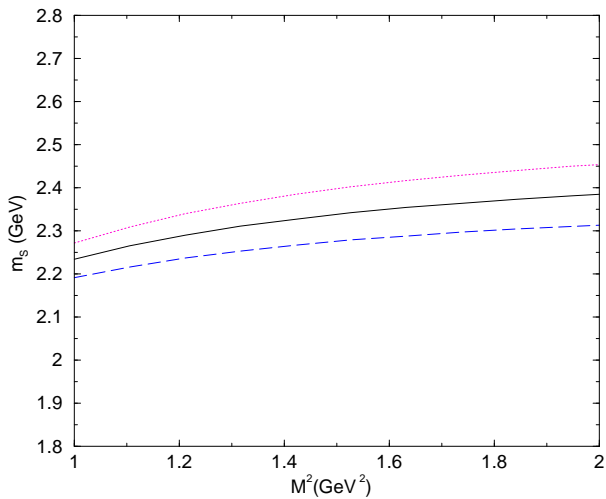,width=8cm}}
%\vspace{-1cm}
\caption{\small{The $D_{0}^{(1s)}$ mass as a function of the Borel mass 
for different
values of the continuum threshold.  Dashed line: $\sqrt{s_0}=2.6\GeV$; solid
line: $\sqrt{s_0}=2.7\GeV$; dotted line: $\sqrt{s_0}=2.8\GeV$.}}
\label{fig5}
\end{figure}

Comparing these two figures we see that 
the mass of $D_{0}^{(0s)}$ is around $100\MeV$ smaller than the others, since
the $D_{0}^{(1s)}$ and $D_{0}^{(2s)}$
resonance masses  are basicaly degenerated \cite{blmnn}. While
it is natural to expect that the inclusion of a strange quark would increase
the resonance mass by around the strange quark mass (as was the case when 
one goes from $D_{0}^{(0s)}$ to $D_{0}^{(1s)}$), it is really interesting
to observe that this does not happen when one goes from $D_{0}^{(1s)}$
to $D_{0}^{(2s)}$. In terms of the OPE contributions, we can trace this
behavior to the fact that the quark condensate term is smaller in
$D_{0}^{(2s)}$ than in $D_{0}^{(1s)}$ (due to the change from $m_c\qq$ to 
$m_c\ss$), however the inclusion of the term
proportional to $m_sm_c$ (which is not present in $D_{0}^{(1s)}$), 
compensates this decrease.

Considering the variations on the quark masses and on the continuum 
threshold discussed above, in the Borel window considered here our results 
for the ressonance masses are given in Table I.
\begin{center}
\small{{\bf Table I:} Numerical results for the resonance masses}
\\
\vskip3mm

\begin{tabular}{|c|c|c|c|}  \hline
resonance & $D_{0}^{(0s)}$  & $D_{0}^{(1s)}$ & $D_{0}^{(2s)}$   \\
\hline
mass (GeV) & $2.22\pm0.15$ &$2.32\pm0.13$  & $2.31\pm0.14$\\
\hline
\end{tabular}\end{center}

Comparing the results in Table I with the resonance masses given by
BABAR, BELLE and FOCUS: $D_{sJ}^+(2317)$,
$D_{0}^0(2308)$ and $D_{0}^{0,+}(2405)$, we see that we can 
identify the four-quark states represented by $D_{0}^{(1s)}$ and 
$D_{0}^{(2s)}$ with the BABAR and BELLE resonances respectively. However,
we do not find a four-quark state whose mass is compatible with the
FOCUS resonances, $D_{0}^{0,+}(2405)$. Therefore, we associate the
FOCUS resonances, $D_{0}^{0,+}(2405)$,
with a scalar $c\bar{q}$ state, since its mass is completly in agreement
with the predictions of the quark model in ref.~\cite{god}. It is also 
interesting to point out that a mass of about $2.4~\GeV$ is also compatible
with the the QCD sum rule calculation for a $c\bar{q}$ scalar meson 
\cite{ht}.

One can argue that while a pole approximation is justified for
the very narrow BABAR resonance, this may not be the case for the
rather broad BELLE and FOCUS resonances. To check if the width of the
resonances could modify the pattern observed in the masses of the
four-quark states, in ref.\cite{blmnn} the phenomenological side of the
sum rule, in Eq.~(\ref{sr}), was modified through the introduction of a 
Breit-Wigner-type resonance form \cite{do02}. It was shown that the best 
agreement 
between the right-hand and left-hand sides of the sum rule
is obtained for $m_S\sim2.2~\GeV$. Therefore our conclusion is that
the inclusion of the width does not change the value of the mass obtained for 
the resonance.

Besides de masses, another important point to understand the nature of the
charmed meson states is their corresponding decay width. One can ask how,
in the present approach, it is possible to obtain a extremely narrow width
for $D_{sJ}^+(2317)$, while the $D_{0}^{0}(2308)$ state remain fairly wide?
To compute the decay width of the hadronic decay ${D_0^{(1s)}}\to D_s^+\pi^0$, 
for example, one has to study the three-point function
\beqa
T_{\mu}(p,\pli,q)=\int d^4x~d^4y~e^{i.\pli.x}~e^{iq.y}
\nn\\
\times
\lag0|T\{
j_{D_s}(x)j_{5\mu}^{\pi}(y)j_{1}^\dagger(0)\}|0\rag,
\lb{3point}
\enqa
where $p=\pli+q$, and the currents for the two pseudoscalar mesons in the 
vertex are
\beqa
j_{5\mu}^{\pi}&=&{1\over\sqrt{2}}(\bar{u}_a\gamma_\mu\gamma_5u_a-\bar{d}_a
\gamma_\mu\gamma_5d_a),\nn\\
j_{D_s}&=&i\bar{s}_a\gamma_5c_a.
\lb{pseu}
\enqa

In the phenomenological side, the three-point function in Eq.~(\ref{3point})
is related with the vertex coupling constant, $g_{D_0^{(1s)}D_s\pi}$, which
is related with the decay width through the relation:
\beqa
\Gamma(D_0^{(1s)}\rightarrow D_s\pi)=
\nn\\
{1\over 16\pi m_{S_1}^3}g_{D_0^{(1s)}
D_s\pi}^2\sqrt{\la(m_{S_1}^2,m_{D_s}^2,m_{\pi}^2)},
\lb{decay}
\enqa
where $\la(a,b,c)=a^2+b^2+c^2-2ab-2ac-2bc$.

For the light scalar mesons considered as diquark- antidiquark states, the 
study of their decay width using the QCD sum rule approach was done in 
ref.\cite{sca}. In Table II we show the results obtained for the different
vertices studied in ref.~\cite{sca}, as well as the experimental values.
\begin{center}
\small{{\bf Table II:} Numerical results for the coupling constants}
\\
\vskip3mm

\begin{tabular}{|c|c|c|}  \hline
vertex & $g(\GeV)$  & $g^{exp}(\GeV)$ \\
\hline
$\sigma\pi^+\pi^-$ & $3.1\pm0.5$ &$2.6\pm0.2$ \\
\hline
$\kappa K^+\pi^-$ & $3.6\pm0.3$ &$4.5\pm0.4$ \\
\hline
$f_0 K^+K^-$ & $1.6\pm0.1$ & \\
\hline
$f_0 \pi^+\pi^-$ & $0.47\pm0.05$ & $1.6\pm0.8$ \\
\hline
\end{tabular}\end{center}

From Table II we see that, although not exactly 
in between the experimental error bars, the hadronic couplings determined
from the QCD sum rule calculation are
consistent with existing experimental data. The biggest discrepancy is for
$g_{f_0\pi^+\pi^-}$ and this can be understood since the $f_0\to\pi^+\pi^-$ 
decay is mediated by one gluon exchange and, therefore,
probably in this case $\alpha_s$ corrections could play an important role.
In the case of the decay
$f_0(a_0)\rightarrow K^+K^-$, the coupling can not be experimentally measured
due to the lack of phase space. 

In the case of the decay ${D_0^{(1s)}}\to D_s^+\pi^0$, for an isoscalar
${D_0^{(1s)}}$, in the QCD sum rule approach one only gets a result different 
from zero for the coupling constant, if one allows a break in the $SU(2)$
symmetry. In this case, the coupling is proportional to the difference
of the masses of the $u$ and $d$ quarks, and the difference of the $u$ and 
$d$ quark condensates. In a preliminary calculation we got
\beq
g_{D_0^{(1s)}D_s\pi}\sim0.06\GeV,
\enq
which gives a decay width $\Gamma(D_0^{(1s)}\rightarrow D_s\pi)\sim8$ KeV.
It is important to notice that, if we have used a isovector current for the
$D_0^{(1s)}$ state instead of an isoscalar current, we would get
$\Gamma(D_0^{(1s)}\rightarrow D_s\pi)\sim260~\MeV$. Therefore, it seems 
possible, in this four-quark scenario, to obtain a extremely narrow width
for $D_{sJ}^+(2317)$, while the $D_{0}^{0}(2308)$ state remain fairly wide.

In conclusion, we have presented a QCD sum rule study of the charmed scalar 
mesons 
considered as diquark-antidiquark states. We found that the masses
of the BABAR,  $D_{sJ}^+(2317)$, and BELLE, $D_{0}^0(2308)$, resonances
can be reproduced by the four-quark states $(cq)(\bar{q}\bar{s})$
and $(cs)(\bar{u}\bar{s})$ respectively. However, the mass of the FOCUS
resonance, $D_{0}^{0,+}(2405)$ can not be reproduced in the four-quark state
picture considered here. Therefore, we interpret it as a  normal
$c\bar{q}$ state, since its mass is in complete  agreement
with the predictions of the quark model in ref.~\cite{god}. We also obtain 
a mass of $\sim 2.2~\GeV$ for
a four-quark scalar state $(cq)(\bar{u}\bar{d})$ which was not yet
observed, and that should be also rather broad.

\end{document}